\documentclass[12pt]{article}

\usepackage{a4wide}
\usepackage[pdftex,usenames,dvipsnames]{color}
\usepackage{graphics}
\usepackage{amsfonts, amssymb}

\newcommand{\nit}{\noindent}

\newcommand{\np}{\newpage}
\newcommand{\dsp}{\displaystyle}
\newcommand{\vs}[1]{\vspace{#1 ex}}
\newcommand{\hs}[1]{\hspace{#1 em}}
\newcommand{\bfr}{\begin{flushright}}
\newcommand{\efr}{\end{flushright}}
\newcommand{\bc}{\begin{center}}
\newcommand{\ec}{\end{center}}
\newcommand{\ben}{\begin{enumerate}}
\newcommand{\een}{\end{enumerate}}

\newcommand{\be}{\begin{equation}}
\newcommand{\ee}{\end{equation}}
\newcommand{\ba}{\begin{array}}
\newcommand{\ea}{\end{array}}
\newcommand{\ct}{\cite}
\newcommand{\bit}{\bibitem}
\newcommand{\dd}[2]{\frac{\partial{#1}}{\partial{#2}}}
\newcommand{\ag}{\alpha}
\newcommand{\bg}{\beta}
\newcommand{\gam}{\gamma}
\newcommand{\del}{\delta}
\newcommand{\eps}{\epsilon}

\newcommand{\thg}{\theta}
\newcommand{\kg}{\kappa}
\newcommand{\lb}{\lambda}
\newcommand{\sg}{\sigma}
\newcommand{\rg}{\rho}

\newcommand{\fg}{\phi}
\newcommand{\vf}{\varphi}
\newcommand{\og}{\omega}
\newcommand{\Gam}{\Gamma}
\newcommand{\Del}{\Delta}
\newcommand{\Fg}{\Phi}
\newcommand{\Sg}{\Sigma}
\newcommand{\Og}{\Omega}
\newcommand{\Lb}{\Lambda}

\newcommand{\cD}{{\cal D}}

\newcommand{\cH}{{\cal H}}

\newcommand{\cP}{{\cal P}}

\newcommand{\lh}{\left(}
\newcommand{\rh}{\right)}
\newcommand{\ld}{\left.}
\newcommand{\rd}{\right.}

\newcommand{\nb}{\nabla}

\newcommand{\ctg}{\mbox{\,cotan\,}}
\newcommand{\cotanh}{\mbox{\,cotanh\,}}

\newcommand{\der}{\partial}

\begin{document}

\pagestyle{empty}

\bfr
{Nikhef/2015-006}
\efr

\bc
{\bf {\large Canonical $D = 1$ supergravity framework for FLRW cosmology}} 
\vs{5}

{\large M.P.\ Bogers}$^a$
\vs{2}

Leiden University, Leiden NL 
\vs{2}

and
\vs{2}

{\large J.W.\ van Holten}$^b$
\vs{2}

Leiden University, Leiden \\
and NIKHEF, Amsterdam NL
\vs{3} 

\today
\ec
\vs{7}

\nit
{\small
{\bf Abstract}\\ 
We construct an extension of standard flat FLRW cosmology with matter, possessing local $D = 1$, $N=1$ 
proper-time supersymmetry. The fundamental equation for the resulting mini-superspace models of quantum 
universes is a Dirac-like analogue of the Friedmann and Wheeler-DeWitt equations. We provide solutions of 
this equation for specific matter configurations based on the supersymmetric $O(3)$ and $O(2,1)$ 
$\sg$-models. It turns out that in the compact model the volume rate of growth of the universe is quantized 
and non-vanishing due to the zero-point energy of the scalar fields. In the non-compact model the spectrum of 
the growth rates is continuous but subject to an uncertainty relation involving the scale and the growth factor.
}
\vfill

\footnoterule 
\nit
{\footnotesize 
$^a$ e-mail: bogers@lorentz.leidenuniv.nl  \\
$^b$ e-mail: v.holten@nikhef.nl }

\np

\pagestyle{plain}
\pagenumbering{arabic}

\section{Introduction} 

Supersymmetry plays a role in quantum gravity on various levels. In supergravity it is part of 
the space-time symmetries, as a grading of diffeomorphism invariance, implying the existence 
of interacting spin-3/2 fermions. In superstring theory it is part of the world-sheet symmetries and
as such it allows for the GSO projection eliminating tachyons from the physical spectrum of 
states. In the canonical formulation of supergravity local supersymmetry imposes a fermionic 
first-class constraint, acting effectively as a square root of the Wheeler-DeWitt equation 
\ct{teitelboim1977}. As this fermionic constraint is linear in the bosonic momenta the analysis 
of the Hamilton-Jacobi equation is considerably simplified \ct{kiefer-luck-moniz2005}. 

The simplifications introduced by supersymmetry can be made particularly clear in the context of 
cosmology. The standard evolution equations for a homogeneous and isotropic universe can be 
cast in the form of a simple hamiltonian system with a finite number of degrees of freedom. This 
system can also include a set of homogeneous scalar fields, thus defining a variety of 
mini-superspace models \ct{hartle-hawking1983,vhrk2013}. By embedding these solutions 
in $D = 4$, $N = 1$ supergravity supersymmetric versions of mini-superspace models are
obtained \ct{cheng-death-moniz,death1996}. 

In this paper we discuss an alternative construction, extending mini-superspace with fermionic 
variables using the supermultiplet calculus developed in \ct{jwvh1995} to create mini-superspace 
models with an intrinsic $D = 1$ supersymmetry. In such a context the quantum version 
of the fermionic first-class constraint takes the form of a standard Dirac equation on some real or 
complex manifold \ct{davis-macfarlane-popat-vholten1984}. Although this construction is not 
a direct reduction of $D = 4$ supergravity, it does possess a local (time-dependent) supersymmetry 
which squares to $D = 1$ time reparametrizations. Therefore this supersymmetry has the algebraic 
features of a reduction, and models obtained by dimensional reduction from $3+1$ to $0+1$ 
dimensions preserving one supersymmetry should form a subclass of the models constructed here.

This paper is organized as follows. We review the construction of bosonic and supersymmetric 
mini-superspace models in sects. 2 and 3, respectively. In sect.\ 4 we discuss their canonical structure, 
followed by the derivation of a supersymmetric Hamilton-Jacobi formalism in sect.\ 5. In sect.\ 6 we
perform the canonical quantization and derive the Dirac equation expressing the fermionic 
first-class constraint. In sects.\ 7 and 8 we apply the general formalism to construct and solve the 
cosmological versions of the supersymmetric massless compact $O(3)$ and non-compact 
$O(2,1)$ non-linear $\sg$-models. The scalar fields in these models parametrize the K\"{a}hler 
manifolds $SO(3)/SO(2) \sim SU(2)/U(1)$ and $SO(2,1)/SO(2) \sim SU(1,1)/U(1)$, respectively 
\ct{davis-mf-vh1982}. We find that in the compact model the momentum of the gravitational 
degree of freedom is quantized, which amounts to a quantization condition on the rate of 
volume expansion of the universe. Finally in sect.\ 9 we present a brief discussion of the effect 
of scalar mass terms on the solutions. Sect.\ 10 summarizes the results.

\section{Homogeneous and isotropic space-time \label{s.1}}

The classical dynamics of an isotropic and homogeneous universe is described by the FLRW-reduction 
of the Einstein equation. In this setting the evolution of the universe is described in terms of the scale 
parameter $a(t) \geq 0$ and the line element
\be
ds^2 = - N^2 dt^2 + a^2 g^*_{ij} dx^i dx^j.
\label{1.1}
\ee
Here $N(t)$ is the lapse-function implementing time-reparametrization invariance, and $g^*_{ij}$ is the 
time-independent comoving metric of 3-dimensional space-like hypersurfaces; owing to the isotropy and 
homogeneity of the universe these hypersurfaces must be either flat, or spherical, or hyperbolic. For 
such geometries the Einstein action per unit comoving volume then reduces to
\be
A = \int dt \lh - \frac{3}{N} a \dot{a}^2 + N (3 ka - \Lb a^3) \rh,
\label{1.2}
\ee
where $k$ is the 3-dimensional curvature constant with values 0 or $\pm 1$ and $\Lb$ is the cosmological 
constant. It is convenient to rewrite the cosmological model by defining an unconstrained scale parameter 
$\vf^0$ and a new time variable $\tau$ by
\be
a = e^{\vf^0/\sqrt{6}}, \hs{2} dt = e^{\frac{1}{2} \sqrt{6}\, \vf^0} d\tau.
\label{1.4}
\ee
Furthermore, the pure FLRW cosmology can be extended with a set of isotropic and homogenous scalar 
fields $\vf^a(t)$, $a = 1, ..., r$. If these scalar fields take values on a real manifold with metric $G_{ab}[\vf]$ 
the action (\ref{1.2}) generalizes to 
\be
A_G = \int d\tau N \left[ - \frac{1}{2} \lh \cD \vf^0 \rh^2 + \frac{1}{2}\, G_{ab} \cD \vf^a \cD \vf^b 
 - U[\vf^0, \vf^a] \right],
\label{1.5}
\ee
where the reparametrization-invariant time derivative is defined by 
\be
\cD = \frac{1}{N} \frac{d}{d\tau},
\label{1.3}
\ee
and the generalized potential $U$ is given by
\be
U[\vf^0, \vf^a] = e^{\sqrt{6}\, \vf^0} V[\vf^a] - 3 k e^{\sqrt{8/3}\, \vf^0}.
\label{1.6}
\ee
Here $V[\vf^a]$ is the scalar potential, taken to include the cosmological constant $\Lb$; the other term 
vanishes in flat space, when $k = 0$. Note that in this formulation the reparametrization-invariant Hubble 
parameter $\cH$ satisfies 
\be
a^3 \cH = \frac{da^3}{3N dt} = \frac{1}{\sqrt{6}}\, \cD \vf^0.
\label{1.7}
\ee 
Thus $\cD \vf^0$ represents the cosmic volume rate of change. As observed by many authors, the action 
(\ref{1.5}) takes the form of a particular version of relativistic particle dynamics in $r$ space dimensions. 
The hamiltonian structure of the particular formulation (\ref{1.5})  of the models was investigated previously 
in ref.\ \ct{vhrk2013}.

\section{$D = 1$ supergravity actions \label{s.2}}

A framework to construct relativistic particle dynamics with world-line supersymmetry in external fields 
is provided by $D = 1$, $N = 1$ supergravity \ct{jwvh1995}. It describes pseudo-classical models of 
spinning-particle dynamics, and quantization of these models culminates in a Dirac equation 
for the wave function. Applying this construction to the cosmological models (\ref{1.5}) we obtain a 
supersymmetric extension of FLRW cosmology. In this context local world-line supersymmetry is a 
graded extension of the group of time-reparametrizations, and as such a subgroup of the local 
supersymmetry of $D = 4$, $N = 1$ supergravity. 

Following ref.\ \ct{jwvh1995} local world-line supersymmetry is realized by introduction of a gauge 
multiplet of world-line one-forms with components $(N, \chi)$, where $N$ is the lapse function and 
$\chi$ a Grassmann variable representing the world-line gravitino. Under local supersymmetry with 
anti-commuting parameter $\eps(\tau)$ we take these variables to transform as
\be
\del N = - 2 i N \eps \chi, \hs{2} \del \chi = \cD \eps = \frac{1}{N} \frac{d\eps}{d\tau}.
\label{2.1}
\ee
The matter fields are contained in scalar supermultiplets $(\vf, \psi)$ with transformation rules
\be
\del \vf = - i \eps \psi, \hs{2} \del \psi =  \lh \cD \vf + i \chi \psi \rh \eps.
\label{2.2}
\ee
For the construction of general invariant actions with potential terms we also need fermionic 
supermultiplets $(\eta, f)$, with $\eta$ a Grassmann variable and $f$ an auxiliary scalar. 
These components transform under supersymmetry as 
\be
\del \eta = f \eps, \hs{2} \del f =  i \lh \cD \eta - f \chi \rh \eps.
\label{2.3}
\ee
The supersymmetric extension of the action $A_G$ in (\ref{1.5}) then takes the form
\be
A_S = A_{kin} + A_{pot},
\label{2.4}
\ee
with
\be
A_{kin} = \int d\tau N \lh \frac{1}{2}\, g_{\mu\nu} \cD \vf^{\mu} \cD \vf^{\nu} + \frac{i}{2}\, g_{\mu\nu} \psi^{\mu} 
 \cD \psi^{\nu} + \frac{i}{2}\, \cD \vf^{\lb} g_{\lb\mu,\nu} \psi^{\mu} \psi^{\nu} - i g_{\mu\nu} \cD \vf^{\mu} \psi^{\nu} \chi \rh,
\label{2.5}
\ee
where 
\be
g_{\mu\nu} = \lh \ba{cc} -1 & 0 \\
                                          0  & G_{ab} \ea \rh.
\label{2.5.1}
\ee
The most straightforward way to add potential terms in flat space $(k = 0)$ is by introduction of $s$ fermionic multiplets
$(\eta_i, f_i)$ with action
\be
A_{pot} = \int d\tau N \lh \frac{i}{2}\, \eta_i \cD \eta_i + \frac{1}{2}\, f_i^2 - W_i (f_i + i \eta_i \chi) 
 + i \eta_i W_{i,\mu} \psi^{\mu} \rh,
\label{2.6}
\ee
with $W[\vf]$ a superpotential such that
\be
\frac{1}{2}\, W_i^2 = U = e^{\sqrt{6}\, \vf^0} V[\vf^a].
\label{2.7}
\ee
More complicated situations are possible when the fermionic variables $\eta_i$ have non-standard 
(e.g., $\vf$-dependent) kinetic terms \ct{jwvh1995}. As a special case, for non-flat universes the additional 
scale-factor dependent term in the potential can be constructed by introducing a second fermionic supermultiplet 
$(\lb, h)$ and adding terms
\be
A_k = k \int d\tau N \lh \frac{i}{2} \lb \cD \lb + \frac{1}{2}\, h^2 - \sqrt{6} e^{\sqrt{2/3}\, \vf^0} (h + i \lb \chi)
 - 2 i e^{\sqrt{2/3}\, \vf^0} \psi^0 \lb \rh.
\label{2.8}
\ee
However, note that for $k = -1$ the fermionic variable $\lb$ will have a negative-sign kinetic term.
In the following we restrict discussions to the physically most relevant case $k = 0$. 

The equations of motion derived from the action $A_S$ can be written in elegant supercovariant form:
\be
\ba{l}
\dsp{ \nb^2 \vf^{\mu} + \Gam_{\lb\nu}^{\;\;\;\mu}\, \nb \vf^{\lb} \nb \vf^{\nu} =
  \frac{i}{2}\, \psi^{\kg} \psi^{\lb} R_{\kg\lb\nu}^{\;\;\;\;\;\mu} \nb \vf^{\nu} + g^{\mu\nu} W_{i,\nu} W_i
   + i g^{\mu\nu} \eta_i \lh W_{i,\nu} \chi - W_{i,\nu\lb} \psi^{\lb} \rh, }\\
 \\
\dsp{ \nb \psi^{\mu} + \Gam_{\lb\nu}^{\;\;\;\mu}\, \nb \vf^{\lb}\, \psi^{\nu} = g^{\mu\nu} W_{i,\nu} \eta_i, }\\
 \\
\dsp{ \nb \eta_i = - W_{i,\mu} \psi^{\mu},  \hs{2} f_i = W_i[\vf], }\\
\ea
\label{2.9}
\ee
where the supercovariant derivatives are defined by 
\be
\ba{l}
\nb \vf^{\mu} = \cD \vf^{\mu} + i \chi \psi^{\mu}, \hs{2.8} \nb \psi^{\mu} = \cD \psi^{\mu} - \chi \nb \vf^{\mu}, \\
 \\
\nb^2 \vf^{\mu} = \cD \nb \vf^{\mu} + i \chi \nb \psi^{\mu}, \hs{1} \nb \eta_i = \cD \eta_i - W_i \chi.
\ea
\label{2.10}
\ee
In addition there are first-class constraints from the variations w.r.t.\ the non-dynamical gauge components $(N, \chi)$:
\be
\ba{l}
\dsp{ \frac{1}{2}\, g_{\mu\nu} \cD \vf^{\mu} \cD \vf^{\nu} + \frac{1}{2}\, W_i^2 + i \eta_i \lh W_i \chi - W_{i,\mu} \psi^{\mu} \rh 
 = 0, }\\
 \\
 g_{\mu\nu} \cD \vf^{\mu} \psi^{\nu} + W_i \eta_i = 0.
\ea
\label{2.11}
\ee
The first equation is actually the Friedmann equation: 
\be
3 e^{\sqrt{6}\, \vf^0} \cH^2 = \frac{1}{2}\, G_{ab} \cD \vf^a \cD \vf^b + \frac{1}{2}\, W_i^2 + i \eta_i \lh W_i \chi - W_{i,\mu} \psi^{\mu} \rh.
\label{2.12}
\ee
The other constraint is the fermionic superpartner of this equation central to cosmology:
\be
\sqrt{6}\, e^{\sqrt{3/2}\, \vf^0}\cH \psi^0 = G_{ab} \cD \vf^a \psi^b + W_i \eta_i.
\label{2.13}
\ee
These equations are the starting point for the construction of cosmology models  in the following.

\section{Canonical structure \label{s.3}}

The supersymmetric equations of cosmology constructed in sect.\ \ref{s.2} are formulated in lagrangean 
configuration space, with the scale factor and scalar matter fields obeying second-order differential
equations in cosmological time $\tau$. The hamiltonian formalism provides a formulation in 
phase space, where all dynamical variables satisfy first-order differential equations in $\tau$. 
Now the fermionic (Grassmann-odd) variables already satisfy first-order equations in the lagrangean
formulation; therefore the hamiltonian description requires mostly a reformulation of the bosonic 
(Grassmann-even) sector of the theory.

Eq.\ (\ref{2.9}) shows that the scalars $\vf^{\mu}$ satisfy second-order differential equations of motion. 
Therefore we define the canonical momenta
\be
P_{\mu} = \frac{\del A_S}{\del \dot{\vf}^{\mu}} = g_{\mu\nu} \nb \vf^{\nu} 
 + \frac{i}{2}\, g_{\mu\nu,\lb} \psi^{\nu} \psi^{\lb},
\label{3.1}
\ee
and perform a Legendre transformation of the action:
\be
S = \int d\tau N \left[ P_{\mu} \cD \vf^{\mu} + \frac{i}{2}\, g_{\mu\nu} \psi^{\mu} \cD \psi^{\nu} 
 + \frac{i}{2}\, \eta_i \cD \eta_i - H \right],
\label{3.2}
\ee
where the hamiltonian is
\be
\ba{lll}
H & = & \dsp{  \frac{1}{2}\, g^{\mu\nu} \lh P_{\mu} - \frac{i}{2} g_{\mu\kg,\lb} \psi^{\kg} \psi^{\lb} 
 - i g_{\mu\lb} \chi \psi^{\lb} \rh \lh P_{\nu} - \frac{i}{2} g_{\nu\rg,\sg} \psi^{\rg} \psi^{\sg} - i g_{\nu\sg} \chi \psi^{\sg} \rh }\\
 & & \\
 & & \dsp{ +\, \frac{1}{2}\, W_i^2 + i \eta_i \lh W_i \chi - W_{i,\mu} \psi^{\mu} \rh. }
\ea
\label{3.3}
\ee
The action $S$ serves as starting point for the analysis of the canonical structure of the theory.
The action is stationary under variations $\lh \del \vf^{\mu}, \del P_{\mu}, \del \psi^{\mu}, \del \eta \rh$ 
provided
\be
\left[ \ba{c} \dd{H}{x^{\mu}} \\ \\ \dd{H}{P_{\mu}} \\ \\ \dd{H}{\psi^{\mu}} \\ \\ \dd{H}{\eta_i} \ea \right] = 
 \lh \ba{cccc} 0 & - \del_{\mu}^{\nu} & \frac{i}{2}\, g_{\nu\lb,\mu} \psi^{\lb} & 0 \\ \\
                     \del_{\nu}^{\mu} & 0 & 0 & 0 \\ \\ 
                     \frac{i}{2}\, g_{\mu\lb,\nu} \psi^{\lb} & 0 & i g_{\mu\nu} & 0 \\ \\
                     0 & 0 & 0 & i \del_{ij}  \ea \rh 
 \left[ \ba{c} \cD x^{\nu} \\ \\ \cD P_{\nu} \\ \\ \cD \psi^{\nu} \\ \\ \cD \eta_j \ea \right].                   
\label{3.4}
\ee
Upon inverting this equation one gets
\be
\left[ \ba{c} \cD x^{\mu} \\ \\ \cD P_{\mu} \\ \\ \cD \psi^{\mu} \\ \\ \cD \eta_i \ea \right] = 
 \lh \ba{cccc} 0 & \del^{\mu}_{\nu} & 0 & 0 \\ \\ 
                    - \del^{\nu}_{\mu} & - \frac{i}{4}\, g^{\ag\bg} g_{\ag\kg,\mu} g_{\bg\lb,\nu} \psi^{\kg} \psi^{\lb} & 
                                                    \frac{1}{2}\, g_{\kg\lb,\mu} g^{\lb\nu} \psi^{\kg} & 0 \\ \\
                     0 & - \frac{1}{2}\, g^{\mu\lb} g_{\lb\kg, \nu} \psi^{\kg} & - i g^{\mu\nu} & 0 \\ \\
                     0 & 0 & 0 & -i \del_{ij} \ea \rh
                     \left[ \ba{c} \dd{H}{x^{\nu}} \\ \\ \dd{H}{P_{\nu}} \\ \\ \dd{H}{\psi^{\nu}} \\ \\ \dd{H}{\eta_j} \ea \right].
\label{3.5}
\ee
We can now generalize the procedure of ref.\ \ct{jackiw1993} to read off the canonical brackets for bosonic 
and fermionic degrees of freedom:
\be
\ba{ll}
\dsp{ \left\{ \vf^{\mu}, P_{\nu} \right\} = \del^{\mu}_{\nu}, }& 
\dsp{ \left\{ P_{\mu}, P_{\nu} \right\} = - \frac{i}{4}\, g^{\ag\bg} g_{\ag\kg,\mu} g_{\bg\lb,\nu} \psi^{\kg} \psi^{\lb}, }\\
 & \\
\dsp{ \left\{ P_{\mu}, \psi^{\nu} \right\} =\frac{1}{2}\, g_{\kg\lb,\mu} g^{\lb\nu} \psi^{\kg}, }&
\dsp{ \left\{ \psi^{\mu}, \psi^{\nu} \right\} = - i g^{\mu\nu}, \hs{1} \left\{ \eta_i, \eta_j \right\} = - i \del_{ij}, }
\ea
\label{3.6}
\ee
such that the equation of motion for functions $F(\vf, P, \psi, \eta)$ on the physical phase space read
\be
\cD F = \left\{ F, H \right\}.
\label{3.7}
\ee
The hamiltonian formulation of the dynamics can be streamlined further by defining the non-canonical
supercovariant momenta
\be
\cP_{\mu} = P_{\mu} - \frac{i}{2}\, g_{\mu\nu, \lb} \psi^{\nu} \psi^{\lb} = g_{\mu\nu} \nb \vf^{\nu}.
\label{3.8}
\ee
Reparametrizing the phase space in terms of these variables, the brackets (\ref{3.6}) take the simple 
covariant form
\be
\ba{ll}
\left\{ \vf^{\mu}, \cP_{\nu} \right\} = \del_{\mu}^{\nu}, & \left\{ \psi^{\mu}, \psi^{\nu} \right\} = - i g^{\mu\nu}, 
 \hs{2} \left\{ \eta_i, \eta_j \right\} = - i \del_{ij}, \\
 & \\
\left\{ \psi^{\mu}, \cP_{\nu} \right\} = - \Gam_{\nu\lb}^{\;\;\;\mu}\, \psi^{\lb}, &
\dsp{ \left\{ \cP_{\mu}, \cP_{\nu} \right\} = - \frac{i}{2}\, \psi^{\kg} \psi^{\lb} R_{\kg\lb\mu\nu}, }
\ea
\label{3.9}
\ee
where $\Gam_{\nu\lb}^{\;\;\;\mu}$ and $R_{\mu\nu\kg\lb}$ are the Riemann-Christoffel connection and
curvature tensor computed from the metric $g_{\mu\nu}$. In terms of these new momentum variables the
hamiltonian reads 
\be
H = H_0 - i \chi Q,
\label{3.10}
\ee
with
\be
H_0 = \frac{1}{2}\, g^{\mu\nu} \cP_{\mu} \cP_{\nu} + \frac{1}{2}\, W_i^2 + i \eta_i W_{i,\mu} \psi^{\mu}, \hs{2}
Q = \cP_{\mu} \psi^{\mu} + W_i \eta_i.
\label{3.11}
\ee
Clearly, for $W_i = 0$ the Grassmann variables $\eta_i$ become  non-interacting degrees of freedom
and can be taken to vanish. 

So far we have taken the gauge variables $(N, \chi)$ to act as if they represent external fields. Considering 
the variation of the action (\ref{3.2}) w.r.t.\ these fields, it follows that they impose the first-class constraints 
\be
Q = 0, \hs{2} H = H_0 = 0.
\label{3.13}
\ee
Using the covariant brackets (\ref{3.9}) it is straightforward to establish that
\be
\left\{ Q, Q \right\} = - 2 i H_0, \hs{2} \left\{ Q, H_0 \right\} = 0.
\label{3.12}
\ee
These relations showing that $Q$ is the conserved supercharge are fully consistent with the constraints (\ref{3.13}). 
Once the first-class constraints (\ref{3.13}) have been derived, the local gauge transformations for supersymmetry 
and time translations may be used to impose gauge conditions eliminating the gauge degrees of freedom. The most 
convenient choice is usually the unitary gauge $N = 1$, $\chi = 0$. This gauge condition preserves rigid supersymmetry
with constant parameter $\eps$.

\section{Hamilton-Jacobi equations}

The Hamilton-Jacobi formulation of general-relativistic dynamics is widely used both for studying classical 
evolution equations and as a semi-classical approach towards quantum theory 
\ct{salopek1998,kiefer-luck-moniz2005}. The supersymmetric formalism also provides useful insight in the 
structure of solutions for dynamical systems in general \ct{alonso-etal2004, townsend2007}. In the present 
context the action (\ref{3.2}) forms the starting point for deriving Hamilton-Jacobi equations for both boson- 
and fermion degrees of freedom. Consider the change in $S$ under arbitrary variations of the phase-space 
variables $(\vf, P, \psi, \eta)$:
\be
\ba{lll}
\del S & = & \dsp{ \int_1^2 d\tau N \left[ \del \vf^{\mu} \lh - \cD P_{\mu} + \frac{i}{2}\, g_{\nu\lb,\mu} \psi^{\nu} \cD \psi^{\lb}
 - \dd{H}{\vf^{\mu}} \rh + \del P_{\mu} \lh \cD \vf^{\mu} - \dd{H}{P_{\mu}} \rh \rd }\\
 & & \\
 & & \dsp{ +\, \del \psi^{\mu} \lh i g_{\mu\nu} \cD \psi^{\nu} + \frac{i}{2}\, g_{\mu\nu,\lb} \cD \vf^{\lb} \psi^{\nu}
  - \dd{H}{\psi^{\mu}} \rh + \del \eta_i \lh \cD \eta_i - \dd{H}{\eta_i} \rh }\\
 & & \\
 & & \dsp{ \ld +\, \cD \lh \del \vf^{\mu} P_{\mu} - \frac{i}{2}\, \del \psi^{\mu} g_{\mu\nu} \psi^{\nu} 
   - \frac{i}{2}\, \del \eta_i \eta_i \rh \right]. }
\ea
\label{4.1}
\ee
If all these variables satisfy the hamiltonian equations of motion, the on-shell action can be 
identified with a supersymmetric generalization of Hamilton's principal function characterized by 
\be
\ld \del S \right|_{on-shell} = \left[ \del \vf^{\mu} P_{\mu} - \frac{i}{2}\, \del \psi^{\mu} g_{\mu\nu} \psi^{\nu} 
   - \frac{i}{2}\, \del \eta_i \eta_i \right]_1^2.
\label{4.1.1}
\ee
Thus variations of the final end point $(\vf(\tau_2), \psi(\tau_2), \eta(\tau_2))$ of the phase-space trajectory 
produce the generalized Hamilton-Jacobi equations
\be
\ld \dd{S}{\vf^{\mu}} \right|_{on-shell} = P_{\mu}, \hs{2} 
\ld \dd{S}{\psi^{\mu}} \right|_{on-shell} = - \frac{i}{2}\, g_{\mu\nu} \psi^{\nu}, \hs{2}
\ld \dd{S}{\eta_i} \right|_{on-shell} = - \frac{i}{2}\, \eta_i.
\label{4.2}
\ee
At the same time variations of the gauge degrees of freedom do not produce any boundary terms, and
\be
\ld \dd{S}{N} \right|_{on-shell} = 0, \hs{2} \ld \dd{S}{\chi} \right|_{on-shell} = 0,
\label{4.3}
\ee
showing explicitly that the principal function is gauge-independent \ct{vhrk2013}. It is now straightforward 
to write the supersymmetry  constraint (\ref{3.13}) as a Hamilton-Jacobi equation:
\be 
g^{\mu\nu}\, \dd{S}{\vf^{\mu}} \dd{S}{\psi^{\nu}} + W_i \dd{S}{\eta_i} = 0.
\label{4.4}
\ee

\section{Quantum theory}

A quantum theory can be associated with the classical dynamics described by the action (\ref{3.2}), (\ref{3.3}) 
by promoting the phase-space degrees of freedom to operators and the brackets (\ref{3.9}) to (anti-)commutation
relations. However, to construct suitable operator representations in a well-defined state-function space it
is convenient to reparametrize the fermionic sector of the phase-space in terms of local pseudo-euclidean 
tangent-space frames. Thus on the configuration space of the $\vf^{\mu}$ we introduce the $n$-bein field
$e_{\mu}^{\;\,m}$ and its inverse $e^{\mu}_{\;\,m}$, $m = (0,...,r)$, such that the metric $g_{\mu\nu}$ 
can be converted to pseudo-euclidean form $\eta_{mn} =$ diag$(-1, +1, ..., +1)$:
\be
g_{\mu\nu} = \eta_{mn}\, e_{\mu}^{\;\,m} e_{\nu}^{\;\,n}, \hs{2} e_{\mu}^{\;\,m} e_m^{\;\,\nu} = \del_{\mu}^{\nu}.
\label{5.1}
\ee
Now define new fermion variables
\be
\psi^m = \psi^{\mu} e_{\mu}^{\;\,m}.
\label{5.2}
\ee
The brackets (\ref{3.9}) are then modified to \ct{davis-macfarlane-popat-vholten1984}
\be
\ba{ll}
\left\{ \vf^{\mu}, \cP_{\nu} \right\} = \del_{\mu}^{\nu}, & \left\{ \psi^m, \psi^n \right\} = - i \eta^{mn}, 
 \hs{2} \left\{ \eta_i, \eta_j \right\} = - i \del_{ij}, \\
 & \\
\left\{ \psi^m, \cP_{\nu} \right\} =  \og_{\nu\;\;n}^{\;\,m}\, \psi^n, &
\dsp{ \left\{ \cP_{\mu}, \cP_{\nu} \right\} = - \frac{i}{2}\, \psi^m \psi^n R_{mn\mu\nu}, }
\ea
\label{5.3}
\ee
where $\og_{\mu}$ is the spin-connection defined by the properties
\be
e_{\mu\;\; |\nu}^{\;\,m} \equiv e_{\mu\;\, ,\nu}^{\;\,m} - \Gam_{\mu\nu}^{\;\;\;\lb}\, e_{\lb}^{\;\,m} 
 - \og_{\mu\;\;n}^{\;\,m}\, e_{\nu}^{\;\,n} = 0, \hs{2} \og_{\mu mn} = - \og_{\mu nm}.
\label{5.4}
\ee
In the quantum theory we associate an operator with each physical degree of freedom: 
\be
\vf^{\mu} \rightarrow \xi^{\mu},  \hs{2} \cP_{\mu} \rightarrow \pi_{\mu}, \hs{2} 
\psi^m \rightarrow \gam^m/\sqrt{2}, \hs{2} \eta_i \rightarrow \ag_i/\sqrt{2},
\label{5.5}
\ee
with (anti-)commutation relations
\be
\ba{ll}
\left[ \xi^{\mu}, \pi_{\nu} \right]_- =  i \del_{\mu}^{\nu}, & \left[ \gam^m, \gam^n \right]_+ = 2 \eta^{mn}, \hs{2}
\left[ \ag_i, \ag_j \right]_+ = 2 \del_{ij}, \\
 & \\
\left[ \gam^m, \pi_{\mu} \right]_- = i \og_{\mu\;\;n}^{\;\,m}\, \gam^n, & 
\dsp{ \left[ \pi_{\mu}, \pi_{\nu} \right]_- =  \frac{1}{2}\, \sg^{mn} R_{mn\mu\nu}. }
\ea
\label{5.6}
\ee
Several comments are in order. First, the fermionic operators define two Clifford algebras with $r+1$ 
generators $\gam^m$ and $s$ generators $\ag_i$, respectively. By adding the operator relation 
$\eta_i \gam^m = - \gam^m \eta_i$ they can be combined formally in a single $(r+s+1)$-dimensional 
Clifford algebra. The operators 
\be
\sg^{mn} \equiv \frac{1}{4} \left[ \gam^m, \gam^n \right]_- 
\label{5.7}
\ee
define the spinor representation of the $(r+1)$-dimensional Lorentz algebra so$(r,1)$:
\be
\left[ \sg^{mn}, \sg^{kl} \right]_- = \eta^{ml} \sg^{nk} - \eta^{mk} \sg^{nl} - \eta^{nl} \sg^{mk} + \eta^{nk} \sg^{ml}.
\label{5.8}
\ee
Similarly one can construct the spinor representation of the algebra so$(s)$ from the anti-symmetrized products 
of the $\ag_i$. 

Next we define a scalar product w.r.t.\ which the operators are self-adjoint. First note, that there exists
a representation of the Clifford algebra satisfying
\be
\lh \gam_0 \gam^{m} \rh^{\dagger} = \gam_0 \gam^{m}.
\label{5.9}
\ee
Then a scalar product in the space of spinors $\Psi[\xi]$ with the desired properties is
\be
\lh \Fg, \Psi \rh =  \int d^r \xi\, e\, \bar{\Fg} \Psi, \hs{2} \bar{\Fg} = \Fg^{\dagger} \gam_0,
\label{5.10}
\ee
where $e(\xi) = |\det e_{\mu}^{\;\,m}|$. On this space a self-adjoint realization of the momentum operator is
\be
\pi_{\mu} = - \frac{i}{\sqrt{e}}\, D_{\mu} \sqrt{e} 
  \equiv - \frac{i}{\sqrt{e}} \lh \dd{}{\xi^{\mu}} - \frac{1}{2}\, \og_{\mu mn} \sg^{mn} \rh \sqrt{e}.
\label{5.11}
\ee
For the simplest models without potential terms: $\eta_i = W_i = 0$ this provides a complete 
description of the operator space of quantum theory. When potential terms are present, the operators 
$\Sg^A = (\gam^m, \ag_i)$ define a $(r+s+1)$-dimensional Clifford algebra with Lorentz-signature:
\be
\left[ \Sg^A, \Sg^B \right]_+ = 2 \eta^{AB}, \hs{2} \sg^{AB} = \frac{1}{4} \left[ \Sg^A, \Sg^B \right]_-,
\label{5.12}
\ee
and the functions $\Psi[\xi]$ are spinors in a $(r + s + 1)$ dimensional lorentzian space. The momentum 
operator is then to be represented by a covariant derivative in which $\sg^{mn}$ is replaced by the 
corresponding subset of the matrices $\Sg^{AB}$ for $[AB] = [mn]$, as $\og_{\mu ij} = \og_{\mu im} = 0$ 
for the flat dimensions $A = i = (r+2, ... ,r+s+1)$.

Although we now have a function space with a scalar product, carrying an operator representation of the 
commutation relations (\ref{5.6}), we still have to construct the operators corresponding to the hamiltonian 
and supercharge (\ref{3.11}). The issue with this construction is operator ordering, which renders the 
translation from classical to quantum theory non-unique. However, for supersymmetric theories this issue 
can be resolved by requiring the supersymmetry algebra (\ref{3.12}) to hold in terms of anti-commutators:
\be 
\left[ Q, Q \right]_+ = 2 H_0.
\label{6.1}
\ee
For the supercharge we take the symmetric expression
\be
\ba{lll}
\sqrt{2}\, Q & = & \dsp{ \frac{1}{2}\, \lh \gam^m e_m^{\;\mu}\, \pi_{\mu} + \pi_{\mu} e_m^{\;\mu} \gam^m \rh 
 + \ag_i W_i = - i \gam^m e_m^{\;\,\mu} D_{\mu} + \ag_i W_i. }
\ea
\label{6.2}
\ee
Then the hamiltonian becomes
\be
H_0 = Q^2 = - \frac{1}{2}\, D^{\mu} D_{\mu} - \frac{1}{4}\, R + \frac{1}{2}\, W_i^2 - i W_{i,\mu}\, e^{\mu}_{\;\,m} \Sg^{mi},
\label{6.3}
\ee
where as in eq.\ (\ref{5.12}):
\[
\Sg^{mi} = \frac{1}{4} \left[ \gam^m, \ag^i \right]_-.
\]
In the quantum theory the constraints (\ref{3.13}) can now be reformulated as a Dirac equation 
\be
\lh - i \gam^m e_m^{\;\,\mu} D_{\mu} + \ag_i W_i \rh \Psi = 0.
\label{6.4}
\ee

\section{The non-linear $O(3)$ $\sg$-model}

As an application of the above constructions we consider the non-linear $O(3)$ $\sg$-model \ct{davis-mf-vh1982}.
This a theory of $3$ scalar fields parametrizing a sphere $S^2$ with radius $1/2g$:
\be
\Fg_1^2 + \Fg_2^2 + \Fg_3^2  = \frac{1}{4g^2}.
\label{6.5}
\ee
In terms of polar co-ordinates:
\be
\Fg_1 = \frac{1}{2g}\, \sin \thg \cos \vf, \hs{1} \Fg_2 = \frac{1}{2g}\, \sin \thg \sin \vf \hs{1}
\Fg_3 = \frac{1}{2g}\, \cos \thg,
\label{6.6}
\ee
and the scalar-field kinetic term in the action becomes
\be
\frac{1}{8g^2}\, (\cD \thg)^2 + \frac{1}{8g^2}\, \sin^2 \thg\, (\cD \vf)^2.
\label{6.7}
\ee
In this parametrization the $3$-bein and the spin connection are given by
\be
e_{\mu}^{\;m} = \lh \ba{ccc} 1 & 0 & 0 \\
                                                0 & (2g)^{-1} & 0 \\
                                                0 & 0 & (2g)^{-1} \sin \thg \ea \rh, \hs{2}
\og_{\vf mn} = \lh \ba{ccc} 0 & 0 & 0 \\
                                                 0 & 0 & \cos \thg \\
                                                 0 & - \cos \thg & 0 \ea \rh,                                            
\label{6.8}
\ee
whilst $ \og_{0 mn} = \og_{\thg mn} = 0$. Taking $\gam^0 = i \sg_3$, $\gam_{1,2} = \sg_{1,2}$, 
the Dirac operator (\ref{6.2}) for the pure $\sg$-model with $W_i = \ag_i$ = 0 is
\be
- i \gam^m e_m^{\;\,\mu} D_{\mu} = \lh \ba {cc} \der_0 & -2 i g D_- \\ 
   & \\
   - 2 i g D_+ & - \der_0 \ea \rh,
\label{6.9}
\ee
where
\be
D_{\pm} = \der_{\thg} + \frac{1}{2}\, \ctg \thg \pm \frac{i}{\sin \thg} \der_{\vf}.
\label{6.9.1}
\ee
The elementary eigenspinors are of the form \ct{camporesi-higuchi1995}
\be
\Psi_n(\xi^0,\thg,\vf) = c_n e^{2 i g \kg_n \xi^0 + i m \vf} \left[ \ba {c} \fg_{n+}(\thg) \\ \fg_{n-}(\thg) \ea \right],
\label{6.10}
\ee
where $\fg_{\pm}$ are solutions of the eigenvalue equations
\be
D_{\mp}D_{\pm} \fg_{n \pm} = \lh \der_{\thg}^2 + \ctg \thg\, \der_{\thg} - \frac{1}{4} 
 - \frac{m^2 + \frac{1}{4}}{\sin^2 \thg} \pm \frac{m\cos \thg}{\sin^2 \thg} \rh \fg_{n\pm} = - \kg_n^2 \fg_{n\pm}.
\label{6.10.1}
\ee
The regular eigenfunctions for positive integer $m$ are \ct{davis-mf-vh1982,camporesi-higuchi1995}
\be
\fg_{n+}(\thg) =  \sin^{m} \thg\, \sqrt{\frac{1 + \cos \thg}{\sin \thg}}\, P_n^{(m- 1/2, m+1/2)}(\cos \thg), \hs{1} 
 m \geq 1, \hs{2} D_+ \fg_{n+} = - \kg_n \fg_{n-},
\label{6.10.2}
\ee
where $P_n^{(a,b)}$ is the Jacobi polynomial of order $n = 0, 1, 2, ...$. For negative $m$ the roles 
of $\fg_{n \pm}$ are reversed under provided $\kg_n \rightarrow - \kg_n$. In both cases the 
eigenvalues $\kg_n$ are roots of the equation
\be
\kg^2_n = \lh n + |m| + \frac{1}{2} \rh^2, \hs{1} n = 0, 1, 2, ..., \hs{1} |m| \geq 1.
\label{6.11}
\ee
As a result in pure states there is a quantization condition for the cosmic expansion rate, characterized by 
a discrete spectrum of values $\kg_n$ for the volume rate of change. Recalling the classical equivalence 
(\ref{1.7}) and applying Ehrenfest's theorem this implies that in the classical large-volume limit the scale 
factor satisfies
\be
\frac{da^3}{Ndt} = \pm \sqrt{6}\, g \kg_n \hs{1} \stackrel{N = 1}{\longrightarrow} \hs{1} a(t) \sim t^{1/3},
\label{6.13}
\ee
in standard cosmological time $t$ with lapse function $N=1$. 

\section{The non-compact $O(2,1)$ $\sg$-model}

The non-linear $\sg$-model on the sphere has a non-compact counterpart on the 2-sheet hyperboloid 
defined in terms of three scalar fields related by
\be
\Fg_1^2 + \Fg_2^2 - \Fg_3^2  = - \frac{1}{4g^2}.
\label{nc.1}
\ee
An axially symmetric parametrization is 
\be
\Fg_1 = \frac{1}{2g}\, \sinh \thg\, \cos \vf, \hs{1} \Fg_2 = \frac{1}{2g}\, \sinh \thg\, \sin \vf, \hs{1}
\Fg_3 = \frac{1}{2g}\, \cosh \thg,
\label{nc.2}
\ee
with kinetic lagrangean
\be
\frac{1}{8g^2}\, (\cD \thg)^2 + \frac{1}{8g^2}\, \sinh^2 \thg\, (\cD \vf)^2.
\label{nc.3}
\ee
For this model the $3$-bein and the spin connection are given by
\be
e_{\mu}^{\;m} = \lh \ba{ccc} 1 & 0 & 0 \\
                                                0 & (2g)^{-1} & 0 \\
                                                0 & 0 & (2g)^{-1} \sinh \thg \ea \rh, \hs{2}
\og_{\vf mn} = \lh \ba{ccc} 0 & 0 & 0 \\
                                                 0 & 0 & \cosh \thg \\
                                                 0 & - \cosh \thg & 0 \ea \rh,                                            
\label{nc.4}
\ee
whilst $\og_{0 mn} = \og_{\thg mn} = 0$. The Dirac operator now takes the form (\ref{6.9}) with
\be
D_{\pm} = \der_{\thg} + \frac{1}{2}\, \cotanh \thg \pm \frac{i}{\sinh \thg} \der_{\vf}
\label{nc.5}
\ee
Again we can look for elementary solutions of the form
\be
\Psi_{\kg}(\xi^0,\thg,\vf) = c_{\kg} e^{2 i g \kg \xi^0 + i m \vf} 
 \left[ \ba {c} \fg_{\kg +}(\thg) \\ \fg_{\kg -}(\thg) \ea \right].
\label{nc.6}
\ee
There is indeed an infinite number of such solutions for integer $m$, given by
\be
\ba{l}
\dsp{ \fg_{\kg +}(\thg;\kg) = \sinh^{m} \thg \sqrt{\frac{1 + \cosh \thg}{\sinh \thg}}\, 
 F\lh m+\frac{1}{2}+ i \kg, m+\frac{1}{2} - i \kg; m+ \frac{1}{2}, - \sinh^2 \frac{\thg}{2} \rh, }\\
 \\
\dsp{ D_+ \fg_{\kg+} = - \kg \fg_{\kg -}, }
\ea
\label{nc.7}
\ee
where $F(\ag,\bg,\gam;x)$ is the hypergeometric function, and the imaginary part of the parameter arguments 
determines $\kg$. The spectrum of eigenvalues $\kg$ is continuous, and the functions (\ref{nc.7}) are only 
$\del$-function normalizable. Therefore the true normalizable wave functions are wave-packets, superpositions 
of solutions $\Psi_{\kg}$ for a range of values of $\kg$:
\be
\Psi(\xi^0, \thg, \vf) = \int_{-\infty}^{\infty} d\kg\, c(\kg) \Psi_{\kg}(\xi^0,\thg,\vf), \hs{2}
\int_{-\infty}^{\infty} d\kg |c(\kg)|^2 = 1.
\label{nc.9}
\ee
These wave packets have no sharp eigenvalues for the momenta $\pi_{\thg}$, which results by 
standard arguments in an uncertainty relation for canonical pair $(\thg, \pi_{\thg})$. As usual, this translates 
into a `time-energy' uncertainty relation as well, of the type
\be
\Del \vf^0 \Del \pi_0 \gtrsim\, \frac{1}{2},
\label{nc.8}
\ee
where $\Del \vf^0$ represents the relative fluctuations in the volume factor $\sim \Del v/v$ during the period in 
which $\thg$ takes values in the range $\Del \thg$, and $\Del \pi_0$ is the corresponding variation in the volume 
growth rate.

\section{Massive scalars}

Cosmological scalar fields representing true Goldstone bosons, such as those of spontaneously broken $O(3)$, 
are strictly massless. However, if they are pseudo-Goldstone bosons of an approximate symmetry, they can get 
a mass which is presumably small compared to the Planck scale. In that case a non-vanishing potential term $W$ 
is present in eq.\ (\ref{6.4}). In view of eq.\ (\ref{2.7}), if the stability group $O(2)$ is preserved and the 
masses are degenerate this potential takes the form
\be
W = \frac{m}{2g} \sin \thg\, e^{\sqrt{3/2}\, \xi^0}.
\label{7.1}
\ee
The price for introducing this term is the introduction of another Dirac matrix, doubling the number of components
in spinor space; a specific representation is
\be
\gam^0 = \lh \ba{cc} i \sg_3 & 0 \\ 0 & - i \sg_3 \ea \rh, \hs{1}
\gam^{1,2} = \lh \ba{cc} \sg_{1,2} & 0 \\ 0 & -  \sg_{1,2} \ea \rh, \hs{1}
\ag = \lh \ba{cc} 0 & 1_2 \\ 1_2 & 0 \ea \rh.
\label{7.2}
\ee
In this representation the Dirac-operator (\ref{6.4}) becomes
\be
\left[ \ba{cc} {\ba{cc} \der_0 & - 2ig D_- \\ -2ig D_+ & - \der_0 \ea} 
                     & \dsp{ \frac{m}{2g}\, \sin \thg\, e^{\sqrt{3/2}\, \xi^0}\, 1_2 }\\
        & \\
        \dsp{ \frac{m}{2g}\, \sin \thg\, e^{\sqrt{3/2}\, \xi^0}\, 1_2 }& {\ba{cc} -\der_0 & 2ig D_- \\ 2ig D_+ & \der_0 \ea} 
        \ea \right].
\label{7.3}
\ee
Thus the Dirac constraint can be written in terms of a pair of 2-component spinors $(\Psi_+, \Psi_-)$ as
\be
\ba{l}
\dsp{ \lh \ba{cc} \der_0 & - 2ig D_- \\ -2ig D_+ & - \der_0 \ea \rh \Psi_+ = - \frac{m}{2g}\, \sin \thg\, e^{\sqrt{3/2}\, \xi^0} \Psi_-,  }\\
 \\ 
\dsp{  \lh \ba{cc} \der_0 & - 2ig D_- \\ -2ig D_+ & - \der_0 \ea \rh \Psi_- = \frac{m}{2g}\, \sin \thg\, e^{\sqrt{3/2}\, \xi^0} \Psi_+. }
\ea
\label{7.4}
\ee
A perturbative scheme for solving these equations can be set up by writing
\be
\Psi_{\pm} = \Psi^{(0)}_{\pm} + m \Psi^{(1)}_{\pm}, 
\label{7.5}
\ee
where $\Psi^{(0)}$ are solutions of the homogeneous equation for massless fields, 
and expanding the equations (\ref{7.4}) to first order in $m$:
\be
 \lh \ba{cc} 2 g \der_0 & - 4ig^2 D_- \\ -4ig^2 D_+ & - 2g \der_0 \ea \rh \Psi^{(1)}_{\pm} = 
  \mp \sin \thg\, e^{\sqrt{3/2}\, \xi^0} \Psi^{(0)}_{\mp} + {\cal O}(m).
\label{7.6}
\ee
It follows, that
\be
2g \lh \ba{cc} \der_0^2 - 4g^2 D_-D_+ & 0 \\ 0 & \der_0^2 - 4g^2 D_+ D_- \ea \rh \Psi^{(1)}_{\pm} =
 \mp e^{\sqrt{3/2}\, \xi^0} \lh \ba{cc} \sqrt{\frac{3}{2}} \sin \thg & - 2ig \cos \thg \\ 
                                            - 2ig \cos \thg & - \sqrt{\frac{3}{2}}\, \sin \thg \ea \rh \Psi_{\mp}^{(0)}.
\label{7.7}
\ee
For an elementary solution $\Psi_n^{(0)}$ as in eq.\ (\ref{6.10}), the perturbation $\Psi_{\pm}^{(1)}$ takes the form
\be
\Psi_{\pm}^{(1)} = c_n e^{i(2g\kg_n - i \sqrt{3/2}) \xi^0 + im \vf} \left[ \ba{c} \zeta_{n+} \\ \zeta_{n-} \ea \right],
\label{7.8}
\ee
with
\be
\lh \ba{cc} \Og_n^2 + D_-D_+ & 0 \\ 0 & \Og_n^2 + D_+ D_- \ea \rh \left[ \ba{c} \zeta_{n+} \\ \zeta_{n-} \ea \right] = 
 \pm \frac{1}{(2g)^3} \lh \ba{cc} \sqrt{\frac{3}{2}} \sin \thg & - 2ig \cos \thg \\ 
                                            - 2ig \cos \thg & - \sqrt{\frac{3}{2}}\, \sin \thg \ea \rh \left[ \ba{c} \fg_{n+} \\ \fg_{n-} \ea \right];
\label{7.9}
\ee
here
\be 
\Og_n^2 = \lh \kg_n - \frac{i\sqrt{6}}{4g} \rh^2. 
\label{7.10}
\ee
Therefore the wave functions $\zeta_{n\pm}$ can be computed in terms of the $\fg_n$ by inverting the differential 
operators $\Og_n^2 + D_{\mp} D_{\pm}$. This is straightforward. Let $\zeta_{n+}$ be decomposed in the
properly normalized basis:
\be
\zeta_{n+} = \sum_m a_{nm} \fg_{m+}, \hs{2} \lh \fg_{n+}, \fg_{m+} \rh = \del_{nm}.
\label{7.11}
\ee
Then 
\be
\ba{lll}
\lh \Og^2_n + D_- D_+ \rh \zeta_{n+} & = & \dsp{ \sum_m \lh \Og^2_n - \kg_m^2 \rh a_{nm} \fg_{m+} }\\
 & & \\
 & = & \dsp{ \frac{1}{(2g)^3}\, \lh \sqrt{\frac{3}{2}} \sin \thg\, \fg_{n+} - 2ig \cos \thg\, \fg_{n-} \rh. }
\ea
\label{7.13}
\ee
Therefore the coefficients $a_{nm}$ can be computed from eq.\ (\ref{7.9}) as
\be
a_{nm} = \frac{1}{(2g)^3 \lh \Og_n^2 - \kg_m^2 \rh} \left[ \sqrt{\frac{3}{2}} \lh \fg_{n+}, \sin \thg\, \fg_{m+} \rh
 + \frac{2ig}{\kg_m} \lh \fg_{n+}, \cos \thg\, D_+ \fg_{m+} \rh \right].
\label{7.12}
\ee
A similar procedure can be applied to solve for $\zeta_{n-}$. 

\section{Conclusions} 

In this paper we have constructed a reduced supersymmetric extension of the Friedmann equation 
for a homogenous and isotropic universe in the presence of scalar and spinor fields, in terms of $D=1$, 
$N=1$ supergravity. We have constructed the Hamilton-Jacobi equations, and we have performed 
the canonical quantization of these models.  The quantization procedure leads to a Dirac-type equation 
for the spinorial wave function of these universes, generalizing the quantized mini-superspace models 
based on a reduced Wheeler-DeWitt equation. In certain cases, e.g.\ when the scalar fields are 
Goldstone bosons such as in the $O(3)$ and $O(2,1)$ $\sg$-models, the equations can be solved 
analytically. In the compact case the rate of  volume growth is quantized. In contrast, in the non-compact 
models the spectrum of the expansion rate is continuous, but an uncertainty relation for volume 
fluctuations and the growth rate of the universe can be derived. Finally, we have described and 
implemented a perturbative procedure for taking into account mass terms for the matter degrees 
of freedom. 

Our method is entirely based on the $N = 1$, $D = 1$ supergravity formalism. It is likely that 
models obtained by dimensional reduction from an $N = 1$, $D = 4$ supergravity parent theory 
can be rewritten in this formalism, but we have not attempted such a construction here. We leave 
this issue for future investigation.
\vs{3}

\nit
{\bf Acknowledgement} \\
For JWvH the work described in this paper is supported by the Foundation for Fundamental 
Research of Matter (FOM) and by the Lorentz Foundation of the University of Leiden.

\np

\end{document}